\documentclass[twocolumn,showpacs,preprintnumbers]{revtex4}%
\usepackage{amssymb}
\usepackage{amsmath}
\usepackage{graphicx}
\usepackage{dcolumn}
\usepackage{bm}
\usepackage{amsfonts}%
\begin{document}
\title{Off resonance coupling between a cavity mode and an ensemble of driven spins}
\author{Hui Wang}
\affiliation{Andrew and Erna Viterbi Department of Electrical Engineering, Technion, Haifa
32000 Israel}
\author{Sergei Masis}
\affiliation{Andrew and Erna Viterbi Department of Electrical Engineering, Technion, Haifa
32000 Israel}
\author{Roei Levi}
\affiliation{Andrew and Erna Viterbi Department of Electrical Engineering, Technion, Haifa
32000 Israel}
\author{Oleg Shtempluk}
\affiliation{Andrew and Erna Viterbi Department of Electrical Engineering, Technion, Haifa
32000 Israel}
\author{Eyal Buks}
\affiliation{Andrew and Erna Viterbi Department of Electrical Engineering, Technion, Haifa
32000 Israel}
\date{\today }

\begin{abstract}
We study the interaction between a superconducting cavity and a spin ensemble.
The response of a cavity mode is monitored while simultaneously the spins are
driven at a frequency close to their Larmor frequency, which is tuned to a
value much higher than the cavity resonance. We experimentally find that the
effective damping rate of the cavity mode is shifted by the driven spins. The
measured shift in the damping rate is attributed to the retarded response of
the cavity mode to the driven spins. The experimental results are compared
with theoretical predictions and fair agreement is found.

\end{abstract}
\maketitle





\section{Introduction}

Cavity quantum electrodynamics (CQED) \cite{Haroche_24} is the study of the
interaction between matter and photons confined in a cavity. In the
Jaynes-Cummings model \cite{Shore_1195} the matter is described using the
two-level approximation, and only a single cavity mode is taken into account.
The interaction has a relatively large effect on the cavity mode response when
the ratio $E/\hbar\omega_{\mathrm{a}}$ between the energy gap $E$ separating
the two levels and the cavity mode photon energy $\hbar\omega_{\mathrm{a}}$ is
tuned close to unity. Recently, it was experimentally found that the cavity
response exhibits higher order resonances in the nonlinear regime when the
ratio $E/\hbar\omega_{\mathrm{a}}$ is tuned close to an integer value larger
than unity \cite{Buks_033807}.%

\begin{figure}
[ptb]
\begin{center}
\includegraphics[
height=2.3726in,
width=3.4537in
]%
{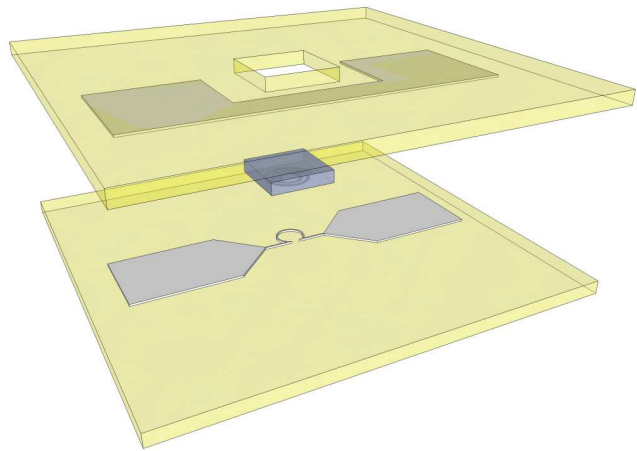}%
\caption{The device is made of two $40\operatorname{mm}\times
40\operatorname{mm}\times0.5\operatorname{mm}$ sapphire wafers carrying the
radio frequency omega resonator, and a $5\operatorname{mm}\times
5\operatorname{mm}\times0.5\operatorname{mm}$ silicon wafer carrying the
microwave spiral resonator. The DPPH powder is placed between the omega
inductor and the spiral. The 3 wafers are vertically shifted in the sketch for
clarity. In the assembled device both the top sapphire wafer and the silicon
wafer are placed directly on top of the bottom sapphire wafer. The three
wafers and a loop antenna are assembled together inside a package made of high
conductivity oxygen free copper. Both omega and spiral resonators are made by
DC-magnetron sputtering of a $200\operatorname{nm}$ thick niobium layer. The
radius of the omega inductor is $500\operatorname{\mu m}$ and the linewidth is
$40\operatorname{\mu m}$. The spiral dimensions are: inner radius
$500\operatorname{\mu m}$, outer radius $580\operatorname{\mu m}$, linewidth
$10\operatorname{\mu m}$ and number of turns $4$. The measured frequency of
the omega (spiral) resonator is $\omega_{\mathrm{a}}/2\pi
=0.173\operatorname{GHz}$ ($\omega_{\mathrm{b}}/2\pi=2.00\operatorname{GHz}$),
whereas the value obtained from numerically simulating the structure is
$0.176\operatorname{GHz}$ ($2.07\operatorname{GHz}$).}%
\label{Fig device}%
\end{center}
\end{figure}

In the current study we explore the case where $E/\hbar\omega_{\mathrm{a}}%
\gg1$ \cite{Ates_724}. This is done by investigating the interaction between
an ensemble of spins and a superconducting cavity mode
\cite{Ghirri_063855,Yap_62,Ghirri_184101}. The energy separation between the
spin energy eigenstates, which is given by $E=\hbar\omega_{\mathrm{L}}$, where
$\omega_{\mathrm{L}}$ is the Larmor frequency, is tuned to a value much higher
than the cavity mode photon energy $\hbar\omega_{\mathrm{a}}$. For this case
the CQED interaction is expected to be negligibly small in the regime of weak
driving. On the other hand, with an intense driving at an angular frequency
close to $\omega_{\mathrm{L}}$ we observe a significant change in the cavity
mode response.

In the current experiment the cavity mode effective damping rate is measured
as a function of the spin driving amplitude and detuning frequency. The
observed shift in the effective damping rate is attributed to the retarded
response of the cavity mode to the driven spins. Related effects of Sisyphus
cooling, amplification, lasing and self-excited oscillation have been
theoretically predicted in other systems having a similar retarded response
\cite{Glenn_195454,Grajcar_612,Voogd_1508_07972,Ella_1210_6902,Ramos_193602}.

\section{Experiment}

Significant change in the response of the measured cavity mode of angular
frequency $\omega_{\mathrm{a}}$ is possible only when intense driving is
applied to the spins. In order to allow sufficiently strong driving, the spin
ensemble is coupled to an additional cavity mode having angular frequency
$\omega_{\mathrm{b}}\gg\omega_{\mathrm{a}}$. When the Larmor frequency
$\omega_{\mathrm{L}}$ is tuned to a value close to $\omega_{\mathrm{b}}$, the
additional cavity mode allows enhancing the spin driving amplitude.

A sketch of the device is seen in Fig. \ref{Fig device}. It is made of two
sapphire wafers and a high resistivity silicon wafer that are attached
together to form a dual band resonator. A radio frequency resonator of angular
frequency $\omega_{\mathrm{a}}$ is constructed by integrating an inductor in
the shape of the greek letter $\Omega$ \cite{Twig_104703} made on the bottom
sapphire wafer, and two capacitors in series, which are formed between the two
sapphire wafers. A square hole is made in the upper sapphire wafer in order to
allow inserting the silicon wafer, which carries a spiral shaped microwave
resonator having angular frequency $\omega_{\mathrm{b}}$
\cite{Maleeva_474,Maleeva_064910}.

\begin{figure}
[ptb]
\begin{center}
\includegraphics[
height=1.7591in,
width=1.6337in
]%
{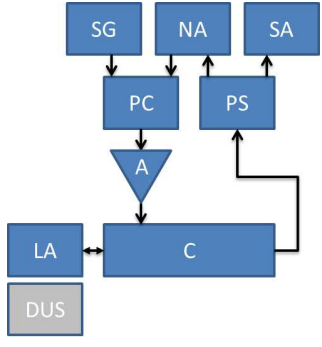}%
\caption{The experimental setup. A power combiner (PC) is employed for
combining the injected signals of a signal generator (SG) and a network
analyzer (NA). The combined injected signal is transmitted through an
amplifier (A) and a coupler (C), and feeds the loop antenna (LA), which is
positioned above the device under study (DUS). The back-reflected signal is
splitted by a power splitter (PS) and measured by both a NA and a spectrum
analyzer (SA).}%
\label{Fig_setup}%
\end{center}
\end{figure}

Both the resonators are designed to be efficiently coupled to the spin
ensemble of diphenylpicrylhydrazyl (DPPH) powder, placed in between them. This
radical, which contains three benzene rings, has a single unpaired electron,
which gives rise to Land\'{e} g-factor of $2.0036$
\cite{Kaplan_1182,Lloyd_1576}. A sketch of the experimental setup is seen in
Fig. \ref{Fig_setup}. A loop antenna is employed for delivering input and
output signals to both resonators.%

\begin{figure}
[ptb]
\begin{center}
\includegraphics[
height=2.6742in,
width=3.4546in
]%
{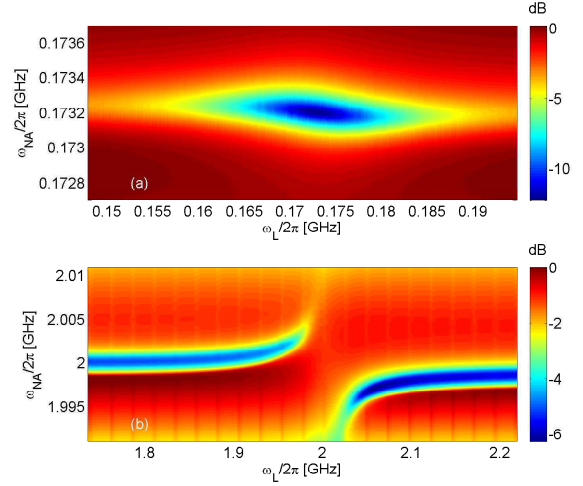}%
\caption{ESR of (a) the omega and (b) the spiral resonators. The color coded
plots display the measured reflectivity coefficient $\left\vert S_{11}%
\right\vert ^{2}$ vs. $\omega_{\mathrm{L}}$ (i.e. vs. static magnetic field)
and the probing frequency $\omega_{\mathrm{NA}}$. Measurements are performed
by a network analyzer at a temperature of $T=3.1\operatorname{K}$, for which
the polarization coefficient $p_{0}$ [see Eq. (\ref{p_0})] is given by
$p_{0}=-1.\,4\times10^{-3}$ ($p_{0}=-1.6\times10^{-2}$) for the omega (spiral)
resonator.}%
\label{Fig ESR}%
\end{center}
\end{figure}

The measured reflectivity near the electron spin resonance (ESR) of the omega
and spiral resonators is seen in Fig. \ref{Fig ESR} (a) and (b), respectively
\cite{Schuster_140501}. Fitting the data with theory [e.g. Eq. (4) of Ref.
\cite{Buks_033807}] allows extracting the value of the coupling coefficient
$g_{\mathrm{a}}$ ($g_{\mathrm{b}}$), which characterizes the interaction
between the spin ensemble and the omega (spiral) resonator, and which is found
to be $g_{\mathrm{a}}=13%
\operatorname{MHz}%
$ ($g_{\mathrm{b}}=83%
\operatorname{MHz}%
$).%

\begin{figure}
[ptb]
\begin{center}
\includegraphics[
height=2.6742in,
width=3.4546in
]%
{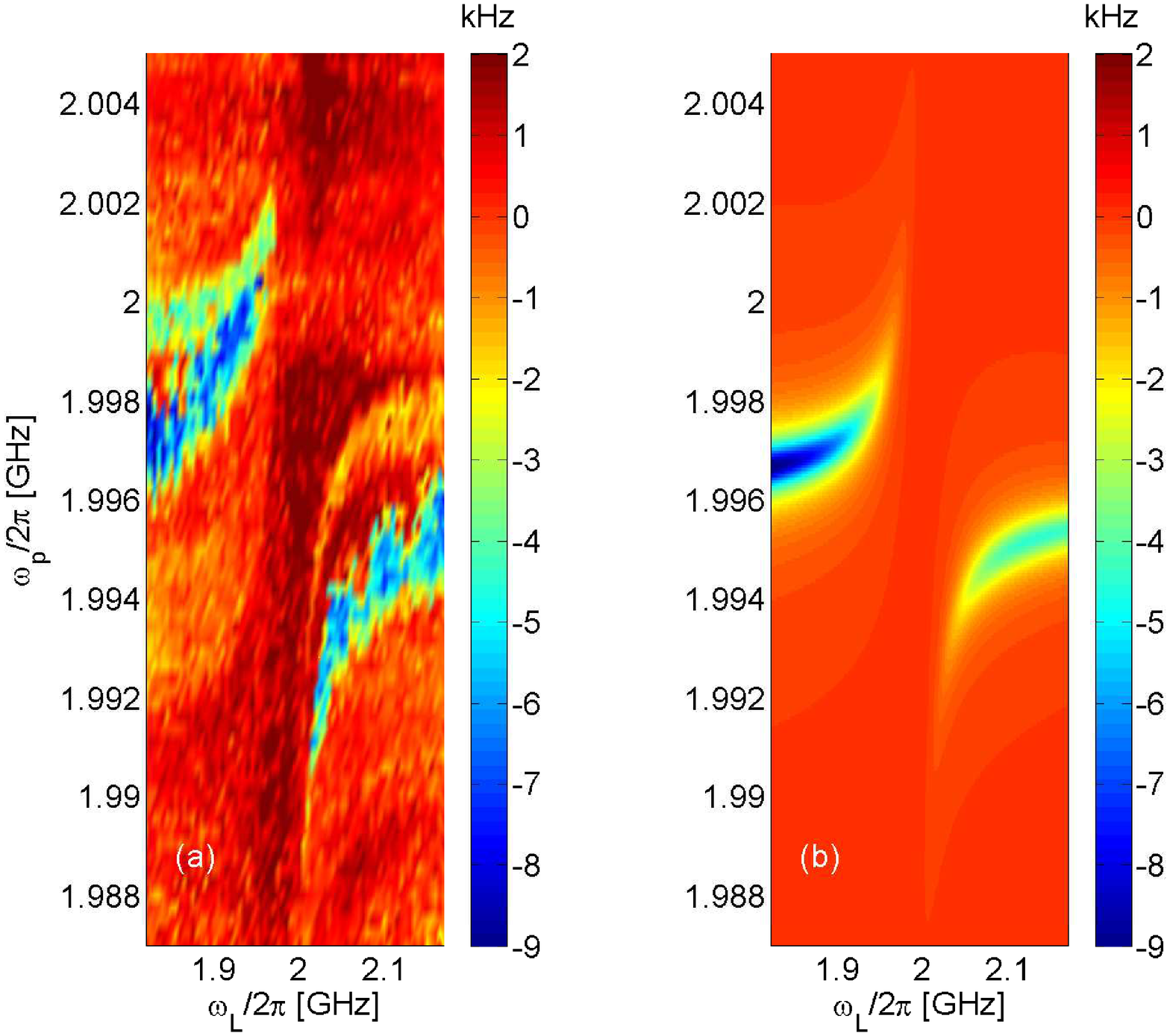}%
\caption{The measured (a) and calculated (b) change in the damping rate
$-\operatorname{Im}\Upsilon_{\mathrm{a}}$ vs. $\omega_{\mathrm{L}}$ and the
pump frequency $\omega_{\mathrm{p}}$. The experimental value is obtained from
the lineshape of the omega resonance. The Larmor frequency $\omega
_{\mathrm{L}}$ is tuned by applying a static magnetic field in a direction
parallel to the wafers. The pump power is set to the value $17{}\mathrm{dBm}$,
which corresponds to a driving amplitude of $\omega_{1}/2\pi
=12\operatorname{MHz}$. The calculated shift (b) is based on Eqs.
(\ref{Upsilon_aL}) and (\ref{Upsilon_ab}). The following parameters are used
in the calculation $\gamma_{\mathrm{b}}=0.4\operatorname{MHz}$ and $\gamma
_{2}=8.3\operatorname{MHz}$ (other parameters are specified above).}%
\label{Fig gamma_ba}%
\end{center}
\end{figure}

The linear response of the decoupled omega resonator is characterized by a
complex angular frequency given by $\omega_{\mathrm{a}}-i\gamma_{\mathrm{a}}$,
where $\gamma_{\mathrm{a}}$ is the mode damping rate. The effect of coupling
on the linear response of the mode can be described in terms of an effective
complex angular frequency $\Omega_{\mathrm{a}}=\omega_{\mathrm{a}}%
-i\gamma_{\mathrm{a}}+\Upsilon_{\mathrm{a}}$, where $\Upsilon_{\mathrm{a}}$
represents the coupling induced frequency shift. The complex angular frequency
$\Omega_{\mathrm{a}}$ can be extracted from the lineshape of the measured
cavity reflectivity vs. frequency curves. The change in the damping rate
$-\operatorname{Im}\Upsilon_{\mathrm{a}}$ is seen in the color-coded plots of
Fig. \ref{Fig gamma_ba} as a function of the Larmor frequency $\omega
_{\mathrm{L}}$ and the spin driving angular frequency $\omega_{\mathrm{p}}$.%

\section{Theory}

To account for the experimental findings, two possible contributions to
$\Upsilon_{\mathrm{a}}$, which is expressed as $\Upsilon_{\mathrm{a}}%
=\Upsilon_{\mathrm{aL}}+\Upsilon_{\mathrm{ab}}$, have been theoretically
estimated. While $\Upsilon_{\mathrm{aL}}$ represents the shift induced by the
coupling to the driven spins, the $\Upsilon_{\mathrm{ab}}$ contribution
originates from the coupling to the driven spiral mode.

A magnetic field having two mutually orthogonal components, a static component
and an alternating one at an angular frequency $\omega_{\mathrm{p}}$, is
applied to the spin ensemble. The amplitude of the static (alternating)
component is $\gamma_{\mathrm{g}}^{-1}\omega_{\mathrm{L}}$ ($\gamma
_{\mathrm{g}}^{-1}\omega_{1}$), where $\gamma_{\mathrm{g}}$ is the electron
spin gyromagnetic ratio. The frequency shift $\Upsilon_{\mathrm{aL}}$ is found
to be given by [see appendix A and Eq. (\ref{Lambda_1= V3})]%
\begin{equation}
\Upsilon_{\mathrm{aL}}=\frac{\frac{\frac{8g_{\mathrm{a}}^{2}\omega_{1}^{2}%
}{\omega_{\mathrm{a}}^{2}\gamma_{2}}\frac{\Delta_{\mathrm{pL}}}{\gamma_{2}%
}\left(  i-\frac{2\gamma_{2}}{\omega_{\mathrm{a}}}\right)  }{1+\frac
{\Delta_{\mathrm{pL}}^{2}}{\gamma_{2}^{2}}+\frac{4\omega_{1}^{2}}{\gamma
_{1}\gamma_{2}}}p_{0}}{\frac{\gamma_{1}}{\omega_{\mathrm{a}}}\left(
\frac{\omega_{\mathrm{R}}^{2}+\eta\omega_{\mathrm{a}}^{2}}{\omega_{\mathrm{a}%
}^{2}}-1\right)  -i\left(  \frac{\omega_{\mathrm{R}}^{2}}{\omega_{\mathrm{a}%
}^{2}}-1\right)  }\;, \label{Upsilon_aL}%
\end{equation}
where $\Delta_{\mathrm{pL}}=\omega_{\mathrm{p}}-\omega_{\mathrm{L}}$ is the
detuning, $\gamma_{1}$ ($\gamma_{2}$) is the longitudinal (transverse) spin
relaxation rate, $p_{0}$ is the spin polarization in thermal equilibrium [see
Eq. (\ref{p_0})], $\omega_{\mathrm{R}}=\sqrt{4\omega_{1}^{2}+\Delta
_{\mathrm{pL}}^{2}}$ is the Rabi frequency of the driven spins and $\eta$ is
given by $\eta=\left(  2\gamma_{2}/\gamma_{1}\right)  \left[  2\omega_{1}%
^{2}\left(  1-\gamma_{1}/\gamma_{2}\right)  /\omega_{\mathrm{a}}^{2}-1\right]
$ [see Eq. (\ref{eta=})]. Note that Eq. (\ref{Upsilon_aL}) is obtained by
assuming that $\left\vert \Delta_{\mathrm{pL}}\right\vert \ll\omega
_{\mathrm{L}}$, $\gamma_{\mathrm{a}}\ll\omega_{\mathrm{a}}$ and $\gamma
_{1},\gamma_{2}\ll\omega_{\mathrm{a}}$.

\begin{figure}
[ptb]
\begin{center}
\includegraphics[
height=2.6862in,
width=3.4546in
]%
{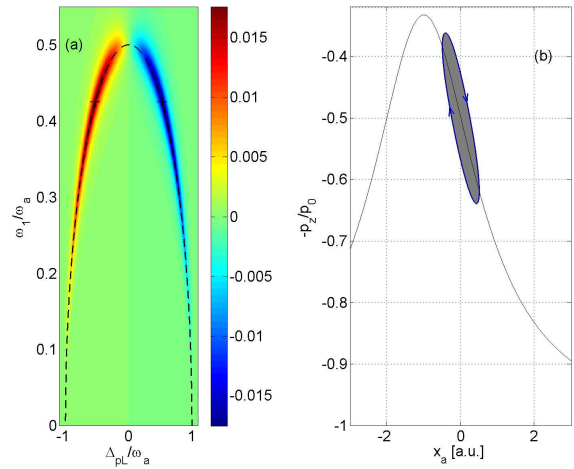}%
\caption{The contribution to cavity mode damping rate $\gamma_{\mathrm{aL}}$
due to coupling to the driven spins. (a) The normalized contribution
$\gamma_{\mathrm{aL}}/\omega_{\mathrm{a}}$ vs. normalized detuning
$\Delta_{\mathrm{pL}}/\omega_{\mathrm{a}}$ and normalized driving amplitude
$\omega_{1}/\omega_{\mathrm{a}}$. The calculation is based on Eq.
(\ref{Upsilon_aL}) with the following assumed parameters $\gamma_{1}%
/\omega_{\mathrm{a}}=2\gamma_{2}/\omega_{\mathrm{a}}=0.05$, $g_{\mathrm{a}%
}/\omega_{\mathrm{a}}=0.1$ and $p_{0}=-0.1$. (b) Normalized spin polarization
$-p_{z}/p_{0}$ vs. cavity mode amplitude $x_{\mathrm{a}}$ for the case of blue
detuned driving. The black solid line represents the steady state normalized
spin polarization $-p_{z0}/p_{0}$. Retardation in the response of the spins to
periodic oscillation of $x_{\mathrm{a}}$ is illustrated by the blue closed
curved.}%
\label{Fig gamma_aL}%
\end{center}
\end{figure}

The real part of $\Upsilon_{\mathrm{aL}}$ is the cavity mode angular frequency
change, induced by the coupling to the driven spins. The imaginary part of
$\Upsilon_{\mathrm{aL}}$ is related to the induced damping rate change
$\gamma_{\mathrm{aL}}$ by $\gamma_{\mathrm{aL}}=-\operatorname{Im}%
\Upsilon_{\mathrm{aL}}$. The color coded plot in Fig. \ref{Fig gamma_aL}(a)
exhibits the dependence of the normalized change in damping rate
$\gamma_{\mathrm{aL}}/\omega_{\mathrm{a}}$ on the normalized detuning
$\Delta_{\mathrm{pL}}/\omega_{\mathrm{a}}$ and the normalized driving
amplitude $\omega_{1}/\omega_{\mathrm{a}}$. When the driving is red detuned ,
i.e. when $\Delta_{\mathrm{pL}}$ is negative, the change in damping rate
$\gamma_{\mathrm{aL}}$ is positive, and consequently mode cooling is expected
to occur \cite{Aspelmeyer_1391}. The opposite behavior occurs with blue
detuning, i.e. when $\Delta_{\mathrm{pL}}$ is positive. For both cases, large
change in the effective cavity mode damping rate occurs near the overlaid
dotted line in Fig. \ref{Fig gamma_aL}(a), along which the Rabi frequency
$\omega_{\mathrm{R}}$ coincides with the cavity mode frequency $\omega
_{\mathrm{a}}$, i.e. $\Delta_{\mathrm{pL}}=\pm\sqrt{\omega_{\mathrm{a}}%
^{2}-4\omega_{1}^{2}}$. This behavior can be attributed to the fact that along
the dotted line, i.e. when $\omega_{\mathrm{R}}=\omega_{\mathrm{a}}$, the
imaginary part of the denominator of Eq. (\ref{Upsilon_aL}) vanishes, and
consequently $\left\vert \Upsilon_{\mathrm{aL}}\right\vert $ peaks. The
largest change in damping rate, which is denoted by $\gamma_{\mathrm{aL,\max}%
}$, can be evaluated by analyzing the expression given by Eq.
(\ref{Upsilon_aL}). In the absence of spin dephasing, i.e. when $\gamma
_{1}/\gamma_{2}=2$, it is found that the largest change, which is given by
$\gamma_{\mathrm{aL,\max}}\simeq0.437\times g_{\mathrm{a}}^{2}p_{0}/\gamma
_{2}$, occurs at the points $\left(  \Delta_{\mathrm{pL}}/\omega_{\mathrm{a}%
},\omega_{1}/\omega_{\mathrm{a}}\right)  \simeq\left(  \pm0.527,0.425\right)
$, which are labeled by crosses in Fig. \ref{Fig gamma_aL}(a). In the current
experiment, however, these points are not accessible since $\omega_{1}%
\ll\omega_{\mathrm{a}}$.%

The underlying mechanism responsible for the change in the effective cavity
mode damping rate that is induced by the coupling to the driven spins can be
described as follows \cite{Aspelmeyer_1391}. As can be seen from the equation
of motion for the cavity mode (\ref{eom a}), the coupling gives rise to a
forcing term acting on the cavity mode, which is proportional to the spin
polarization that is denoted by $p_{z}$. On the other hand, as can be seen
from Eq. (\ref{eom p_+}) below, the same coupling effectively shifts the
Larmor frequency of the spins, and consequently the effective spin driving
detuning, which is given by $\Delta_{\mathrm{pL,eff}}=\Delta_{\mathrm{pL}%
}+g_{\mathrm{a}}x_{\mathrm{a}}$, becomes dependent on the cavity mode
amplitude $x_{\mathrm{a}}$ [see Eq. (\ref{x_a})]. For any fixed value of the
cavity mode amplitude $x_{\mathrm{a}}$ the spin polarization $p_{z}$ in steady
state, which is denoted by $p_{z0}$, can be calculated using Eq. (\ref{p_z0})
below. The dependence of $p_{z0}$ on $x_{\mathrm{a}}$ is demonstrated by the
solid black line in Fig. \ref{Fig gamma_aL}(b) for the case of blue detuned
spin driving. Consider first the adiabatic case, for which it is assumed that
$\omega_{\mathrm{a}}\ll\gamma_{1,2}$. For this case the dynamics of the cavity
mode is assumed to be relatively slow, and consequently the spin polarization
$p_{z}$ is expected to remain very close to the steady state value given by
$p_{z0}$, i.e. to adiabatically follow the $x_{\mathrm{a}}$ dependent
instantaneous steady state value. Therefore, no change in the cavity mode
damping rate is expected in the adiabatic limit. This behavior is consistent
with the fact that $\operatorname{Im}\Upsilon_{\mathrm{aL}}\rightarrow0$ in
the limit $\omega_{\mathrm{a}}/\gamma_{1,2}\rightarrow0$. Note, however, that
the expression given by Eq. (\ref{Upsilon_aL}) is not valid in the adiabatic limit.

Large deviation between the momentary polarization $p_{z}$ and the steady
state value $p_{z0}$ is possible in the non-adiabatic case, where the response
of the spins to the time evolution of the cavity mode becomes retarded. The
closed curve in Fig. \ref{Fig gamma_aL}(b) represents the periodic time
evolution of $p_{z}$ for the case where the cavity mode oscillates at a fixed
amplitude at its resonance frequency around the point $x_{\mathrm{a}}=0$.
Since $p_{z}$ is proportional to the force acting on the cavity mode, the area
colored in gray in Fig. \ref{Fig gamma_aL}(b) is proportional to the net work
done on the cavity mode per cycle. While the area is positive for the case of
blue detuning, which is demonstrated by Fig. \ref{Fig gamma_aL}(b), red
detuning gives rise to negative values, i.e. to energy flow away from the
cavity mode. These affects of energy flow between the cavity mode and the
driven spins give rise to the above discussed change in the effective cavity
mode damping rate.

The frequency shift due to the driven spiral mode is attributed to an
intermode coupling term in the Hamiltonian of the coupled system, which is
assumed to be given by $K\left(  A_{\mathrm{a}}+A_{\mathrm{a}}^{\dag}\right)
\left(  A_{\mathrm{b}}+A_{\mathrm{b}}^{\dag}\right)  ^{2}$, where
$A_{\mathrm{a}}$ ($A_{\mathrm{b}}$) is an annihilation operator of the omega
(spiral) resonator, and $K$ is the intermode coupling coefficient [see
Eq.(\ref{H_ab})]. The contribution $\Upsilon_{\mathrm{ab}}$ is found to be
given by [see appendix B and Eqs. (\ref{CT1}) and (\ref{CT2})]%
\begin{align}
\Upsilon_{\mathrm{ab}}  &  =\frac{4K^{2}\left\vert F_{\mathrm{bf}}\right\vert
^{2}}{\omega_{\mathrm{D}}^{2}+\gamma_{\mathrm{b}}^{2}}\nonumber\\
&  \left(  \frac{\frac{\omega_{\mathrm{D}}}{\gamma_{\mathrm{b}}^{2}}}{\left(
\frac{i\left(  \omega_{\mathrm{a}}-\omega_{\mathrm{D}}\right)  }%
{\gamma_{\mathrm{b}}}-1\right)  \left(  \frac{i\left(  \omega_{\mathrm{a}%
}+\omega_{\mathrm{D}}\right)  }{\gamma_{\mathrm{b}}}-1\right)  }+\frac
{1+\frac{i\gamma_{\mathrm{b}}}{\omega_{\mathrm{s}}}}{\omega_{\mathrm{s}}%
}\right)  \;,\nonumber\\
&  \label{Upsilon_ab}%
\end{align}
where $F_{\mathrm{bf}}$ and $\omega_{\mathrm{D}}$ are the amplitude and
angular frequency detuning, respectively, of the spiral mode driving,
$\omega_{\mathrm{b}}$ and $\gamma_{\mathrm{b}}$ are the spiral mode angular
frequency and damping rate, respectively, and $\omega_{\mathrm{s}}%
=2\omega_{\mathrm{b}}-\omega_{\mathrm{a}}$. Note that when $\gamma
_{\mathrm{b}}\ll\omega_{\mathrm{a}}$ and $\gamma_{\mathrm{b}}\ll
\omega_{\mathrm{s}}$, the first term in the second row of Eq.
(\ref{Upsilon_ab}) becomes negligibly small provided that $\left\vert
\omega_{\mathrm{D}}\right\vert \ll\omega_{\mathrm{a}}^{2}/\omega_{\mathrm{s}}$.

\section{Discussion}

As can be seen from the comparison between Fig. \ref{Fig gamma_ba}(a) and Fig.
\ref{Fig gamma_ba}(b), fair agreement is obtained between data and theory.
Reasonable agreement cannot be obtained unless both contributions
$\Upsilon_{\mathrm{aL}}$ [Eq. (\ref{Upsilon_aL})] and $\Upsilon_{\mathrm{ab}}$
[Eq. (\ref{Upsilon_ab})] are taken into account. The contribution of
$\Upsilon_{\mathrm{ab}}$ is dominated by the second term in the second row of
Eq. (\ref{Upsilon_ab}).

Our results demonstrate the ability to modify the effective damping rate of a
cavity mode by driving spins that are coupled to the mode. Red detuned driving
provides a positive contribution to the damping rate, whereas negative
contribution can be obtained by blue detuned driving. For the former case this
effect can be utilized for cooling down a cavity mode, while the later case of
blue detuning may allow the self excitation of oscillation. Operating close to
the threshold of self-excited oscillation, i.e. close to the point where the
total effective damping vanishes, may be useful for some sensing applications,
since the system is expected to become highly responsive to external
perturbations near the threshold.

As was shown above, relatively large change in the damping rate can be induced
provided that the Rabi frequency $\omega_{\mathrm{R}}$ of the driven spins
becomes comparable to the cavity mode frequency $\omega_{\mathrm{a}}$ (see
Fig. \ref{Fig gamma_aL}). Unfortunately, this region is inaccessible with the
devices that have been investigated in the current experiment. However, in
other CQED systems the condition $\omega_{\mathrm{R}}\simeq\omega_{\mathrm{a}%
}$ can be more easily satisfied. For example, with superconducting CQED
systems both strong \cite{Wallraff_162,Houck_080502,Blais_062320,Koch_042319}%
\ and ultra-strong \cite{Niemczyk_772,Forn_237001} coupling is possible . This
together with the ability to drive a Josephson qubit with Rabi frequencies
high in the radio frequency band, may allow satisfying the condition
$\omega_{\mathrm{R}}\simeq\omega_{\mathrm{a}}$ with a strongly coupled cavity
mode. As was shown above, a large change in cavity mode damping rate, on the
order of $g_{\mathrm{a}}^{2}\left\vert p_{0}\right\vert /\gamma_{2}$, is
possible provided that the region where $\omega_{\mathrm{R}}\simeq
\omega_{\mathrm{a}}$ becomes accessible. For a typical superconducting CQED
system, the damping rate of a decoupled cavity mode is far smaller than
$g_{\mathrm{a}}^{2}\left\vert p_{0}\right\vert /\gamma_{2}$, and thus reaching
this region may allow efficiently cooling down cavity modes by off-resonance
qubit driving.

This work was supported by the Israel Science Foundation, the bi-national
science foundation, the Security Research Foundation in the Technion and the
Russell Berrie Nanotechnology Institute. HW acknowledges support by the Shatz
fellowships and by the Viterbi fellowships.

\appendix

\section{Coupling to driven spins}

Consider an ensemble of spin $1/2$ particles coupled to a cavity mode. The
ensemble is characterized by a longitudinal (spin-lattice) relaxation rate
$\gamma_{1}$ and by a transverse (spin-spin) relaxation rate $\gamma_{2}$. An
external magnetic field is applied, having a component alternating with
angular frequency $\omega_{\mathrm{p}}$, and an orthogonal static component.
The amplitude of the alternating (static) component is $\gamma_{\mathrm{g}%
}^{-1}\omega_{1}$ ($\gamma_{\mathrm{g}}^{-1}\omega_{\mathrm{L}}$), where
$\gamma_{\mathrm{g}}=2\pi\times28.03%
\operatorname{GHz}%
\operatorname{T}%
^{-1}$ is the electron spin gyromagnetic ratio. It is assumed that driving is
applied close to the electron spin resonance, i.e. $\left\vert \Delta
_{\mathrm{pL}}\right\vert \ll\omega_{\mathrm{L}}$, where $\Delta_{\mathrm{pL}%
}=\omega_{\mathrm{p}}-\omega_{\mathrm{L}}$ is the detuning. The cavity mode is
characterized by an angular frequency $\omega_{\mathrm{a}}$ and a damping rate
$\gamma_{\mathrm{a}}$. The coupling between the cavity mode and the spin
ensemble is characterized by a longitudinal coupling coefficient
$g_{\mathrm{a}}$.

\subsection{Equations of motion}

The Hamiltonian of the closed system is taken to be given by%
\begin{align}
\hbar^{-1}\mathcal{H}_{\mathrm{aL}}  &  =\omega_{\mathrm{a}}\left(
A_{\mathrm{a}}^{\dag}A_{\mathrm{a}}+\frac{1}{2}\right)  +\frac{\omega
_{\mathrm{L}}}{2}\Sigma_{z}\nonumber\\
&  +\omega_{1}\left(  e^{-i\omega_{\mathrm{p}}t}\Sigma_{+}+e^{i\omega
_{\mathrm{p}}t}\Sigma_{-}\right) \nonumber\\
&  -g_{\mathrm{a}}\left(  A_{\mathrm{a}}+A_{\mathrm{a}}^{\dag}\right)
\Sigma_{z}\ .\nonumber\\
&  \label{H_0 Sigma}%
\end{align}
The Heisenberg equations of motion are generated according to%
\begin{equation}
\frac{\mathrm{d}O}{\mathrm{d}t}=-i\left[  O,\hbar^{-1}\mathcal{H}%
_{\mathrm{aL}}\right]  \;, \label{Heisenberg}%
\end{equation}
where $O$ is an operator. Using the commutation relations%
\begin{align}
\left[  A_{\mathrm{a}},A_{\mathrm{a}}^{\dagger}\right]   &  =1\;,\\
\left[  \Sigma_{z},\Sigma_{+}\right]   &  =2\Sigma_{+}\;,\\
\left[  \Sigma_{z},\Sigma_{-}\right]   &  =-2\Sigma_{-}\;,\\
\left[  \Sigma_{+},\Sigma_{-}\right]   &  =\Sigma_{z}\;,
\end{align}
one obtains%
\begin{equation}
\frac{\mathrm{d}A_{\mathrm{a}}}{\mathrm{d}t}+i\omega_{\mathrm{a}}%
A_{\mathrm{a}}-ig_{\mathrm{a}}\Sigma_{z}=0\;,
\end{equation}%
\begin{equation}
\frac{\mathrm{d}\Sigma_{+}}{\mathrm{d}t}-i\Omega_{\mathrm{L}}\Sigma
_{+}+i\omega_{1}^{\dag}e^{i\omega_{\mathrm{p}}t}\Sigma_{z}=0\;,
\end{equation}
and%
\begin{equation}
\frac{\mathrm{d}\Sigma_{z}}{\mathrm{d}t}+2i\left(  \Sigma_{+}\omega
_{1}e^{-i\omega_{\mathrm{p}}t}-\Sigma_{-}\omega_{1}^{\dag}e^{i\omega
_{\mathrm{p}}t}\right)  =0\;,
\end{equation}
where%
\begin{equation}
\Omega_{\mathrm{L}}=\omega_{\mathrm{L}}-2g_{\mathrm{a}}\left(  A_{\mathrm{a}%
}+A_{\mathrm{a}}^{\dag}\right)  \;.
\end{equation}

In the next step damping is introduced, and the resultant equations for the
operators $A_{\mathrm{a}}$, $\Sigma_{+}$ and $\Sigma_{z}$ are thermally
averaged. This procedure leads to%
\begin{align}
\frac{\mathrm{d}a}{\mathrm{d}t}+\Theta_{\mathrm{a}}  &  =0\;,\label{eom a}\\
\frac{\mathrm{d}p_{+}}{\mathrm{d}t}+\Theta_{+}  &  =0\;,\label{eom p_+}\\
\frac{\mathrm{d}p_{z}}{\mathrm{d}t}+\Theta_{z}  &  =0\;, \label{eom p_z}%
\end{align}
where%
\begin{align}
a  &  =\left\langle A_{\mathrm{a}}\right\rangle \;,\\
p_{+}  &  =e^{-i\omega_{\mathrm{p}}t}\left\langle \Sigma_{+}\right\rangle
\;,\\
p_{z}  &  =\left\langle \Sigma_{z}\right\rangle =p_{z}\;,
\end{align}
triangle brackets denote thermal averaging, the functions $\Theta_{\mathrm{a}%
}$, $\Theta_{+}$ and $\Theta_{z}$ are given by%
\begin{equation}
\Theta_{\mathrm{a}}=\lambda_{\mathrm{a}}a-ig_{\mathrm{a}}p_{z}\;,
\end{equation}%
\begin{gather}
\Theta_{+}=\left(  i\Delta_{\mathrm{pL}}+\gamma_{2}\right)  p_{+}\nonumber\\
+i\omega_{1}p_{z}+2ig_{\mathrm{a}}\left(  a+a^{\ast}\right)  p_{+}%
\;,\nonumber\\
\end{gather}%
\begin{gather}
\Theta_{z}=\gamma_{1}\left(  p_{z}-p_{0}\right)  +2i\omega_{1}\left(
p_{+}-p_{+}^{\ast}\right)  \;,\nonumber\\
\end{gather}
the cavity eigenvalue $\lambda_{\mathrm{a}}$ is given by $\lambda_{\mathrm{a}%
}=i\omega_{\mathrm{a}}+\gamma_{\mathrm{a}}$, the coefficient%
\begin{equation}
p_{0}=-\tanh\left(  \frac{\hbar\omega_{\mathrm{L}}}{2k_{\mathrm{B}}T}\right)
\ , \label{p_0}%
\end{equation}
is the value of $p_{z}$ in thermal equilibrium in the absence of both driving
and coupling, $k_{\mathrm{B}}$ is the Boltzmann's constant and $T$ is the temperature.

\subsection{The cavity eigenvalue}

The $5\times5$ Jacobian matrix%
\begin{equation}
J=\frac{\partial\left(  \Theta_{\mathrm{a}},\Theta_{\mathrm{a}}^{\ast}%
,\Theta_{+},\Theta_{+}^{\ast},\Theta_{z}\right)  }{\partial\left(  a,a^{\ast
},p_{+},p_{+}^{\ast},p_{z}\right)  }\;,
\end{equation}
can be expressed as $J=J_{0}+g_{\mathrm{a}}V$, where the matrix $J_{0}$ in a
block form is given by%
\begin{equation}
J_{0}=\left(
\begin{tabular}
[c]{c|c}%
$%
\begin{array}
[c]{cc}%
\lambda_{\mathrm{a}} & 0\\
0 & \lambda_{\mathrm{a}}^{\ast}%
\end{array}
$ & $0$\\\hline
$0$ & $J_{\mathrm{L}}$%
\end{tabular}
\ \ \ \ \right)  \;,
\end{equation}
the block $J_{\mathrm{L}}$ is given by%
\begin{equation}
J_{\mathrm{L}}=\left(
\begin{array}
[c]{ccc}%
i\Delta_{\mathrm{pL}}+\gamma_{2} & 0 & i\omega_{1}\\
0 & -i\Delta_{\mathrm{pL}}+\gamma_{2} & -i\omega_{1}\\
2i\omega_{1} & -2i\omega_{1} & \gamma_{1}%
\end{array}
\right)  \ , \label{J_a}%
\end{equation}
the matrix $V$ is given by%
\begin{equation}
V=\left(
\begin{array}
[c]{ccccc}%
0 & 0 & 0 & 0 & -i\\
0 & 0 & 0 & 0 & i\\
2ip_{+} & 2ip_{+} & ix_{\mathrm{a}} & 0 & 0\\
-2ip_{+}^{\ast} & -2ip_{+}^{\ast} & 0 & -ix_{\mathrm{a}} & 0\\
0 & 0 & 0 & 0 & 0
\end{array}
\right)  \;, \label{J V}%
\end{equation}
and%
\begin{equation}
x_{\mathrm{a}}=2\left(  a+a^{\ast}\right)  \;. \label{x_a}%
\end{equation}

Let $\lambda_{1}$, $\lambda_{2}$,$\cdots$ ,$\lambda_{5}$ be the five
eigenvalues of $J=J_{0}+g_{\mathrm{a}}V$. In the limit $g_{\mathrm{a}%
}\rightarrow0$, i.e. when the cavity mode is decoupled from the spins, it is
assumed that $\lambda_{1}\rightarrow\lambda_{\mathrm{a}}$. When $g_{\mathrm{a}%
}$ is sufficiently small the eigenvalue $\lambda_{1}$, which henceforth is
referred to as the cavity eigenvalue, can be calculated using perturbation
theory. For the case of high quality factor (i.e. the case where
$\gamma_{\mathrm{a}}\ll\omega_{\mathrm{a}}$) $\lambda_{1}$ is found to be
given to second order in $g_{\mathrm{a}}$ by%
\begin{equation}
\lambda_{1}=i\omega_{\mathrm{a}}+\gamma_{\mathrm{a}}+g_{\mathrm{a}}%
V_{11}-g_{\mathrm{a}}^{2}\left(  VR\left(  \omega_{\mathrm{a}}\right)
V\right)  _{11}+O\left(  g_{\mathrm{a}}^{3}\right)  \;, \label{lambda_1=}%
\end{equation}
where the $5\times5$ matrix $R\left(  \omega\right)  $ in a block form is
given by%
\begin{equation}
R\left(  \omega^{\prime}\right)  =\left(
\begin{tabular}
[c]{c|c}%
$%
\begin{array}
[c]{cc}%
0 & 0\\
0 & 0
\end{array}
$ & $0$\\\hline
$0$ & $\chi_{\mathrm{L}}\left(  \omega^{\prime}\right)  $%
\end{tabular}
\ \ \ \right)  \;,
\end{equation}
where the $3\times3$ spin susceptibility matrix $\chi_{\mathrm{L}}\left(
\omega^{\prime}\right)  $ is given by%
\begin{equation}
\chi_{\mathrm{L}}\left(  \omega^{\prime}\right)  =\left(  J_{\mathrm{L}%
}-i\omega^{\prime}\right)  ^{-1}\;.
\end{equation}
With the help of Eq. (\ref{J V}) one finds that%
\begin{equation}
\lambda_{1}=i\omega_{\mathrm{a}}+\gamma_{\mathrm{a}}+\Lambda_{1}+O\left(
g_{\mathrm{a}}^{3}\right)  \;,
\end{equation}
where%
\begin{equation}
\Lambda_{1}=2g_{\mathrm{a}}^{2}\left[  p_{+}^{\ast}\left(  \chi_{\mathrm{L}%
}\left(  \omega_{\mathrm{a}}\right)  \right)  _{32}-p_{+}\left(
\chi_{\mathrm{L}}\left(  \omega_{\mathrm{a}}\right)  \right)  _{31}\right]
\;. \label{Lambda_1= V1}%
\end{equation}

The following holds [see Eq. (\ref{J_a})]%
\begin{gather}
\chi_{\mathrm{L}}\left(  \omega_{\mathrm{a}}\right) \nonumber\\
=\frac{1}{D_{\mathrm{L}}}\left(
\begin{array}
[c]{ccc}%
D_{2}D_{3}+2\omega_{1}^{2} & 2\omega_{1}^{2} & -i\omega_{1}D_{2}\\
2\omega_{1}^{2} & D_{1}D_{3}+2\omega_{1}^{2} & i\omega_{1}D_{1}\\
-2i\omega_{1}D_{2} & 2i\omega_{1}D_{1} & D_{1}D_{2}%
\end{array}
\right)  \ ,\nonumber\\
\end{gather}
where%
\begin{align}
D_{1}  &  =i\Delta_{\mathrm{pL}}+\gamma_{2}-i\omega_{\mathrm{a}}%
\;,\label{D_1}\\
D_{2}  &  =-i\Delta_{\mathrm{pL}}+\gamma_{2}-i\omega_{\mathrm{a}%
}\;,\label{D_2}\\
D_{3}  &  =\gamma_{1}-i\omega_{\mathrm{a}}\;,\label{V}\\
D_{\mathrm{L}}  &  =D_{1}D_{2}D_{3}+2\omega_{1}^{2}\left(  D_{1}+D_{2}\right)
\;. \label{D_a}%
\end{align}
The determinant $D_{\mathrm{L}}$ can be expressed as [see Eq. (\ref{D_a})]%
\begin{equation}
\frac{D_{\mathrm{L}}}{\omega_{\mathrm{a}}^{3}}=\frac{\gamma_{1}}%
{\omega_{\mathrm{a}}}\left(  \frac{\Delta_{\mathrm{pL}}^{2}-\omega
_{\mathrm{dR}}^{2}}{\omega_{\mathrm{a}}^{2}}\right)  -i\left(  \frac
{\Delta_{\mathrm{pL}}^{2}-\omega_{\mathrm{dI}}^{2}}{\omega_{\mathrm{a}}^{2}%
}\right)  \;, \label{D_a/omega_C^3}%
\end{equation}
where%
\begin{equation}
\frac{\omega_{\mathrm{dR}}}{\omega_{\mathrm{a}}}=\sqrt{1+\frac{2\gamma_{2}%
}{\gamma_{1}}\left(  1-\frac{2\omega_{1}^{2}}{\omega_{\mathrm{a}}^{2}}\right)
-\frac{\gamma_{2}^{2}}{\omega_{\mathrm{a}}^{2}}}\;, \label{omega_dR}%
\end{equation}
and%
\begin{equation}
\frac{\omega_{\mathrm{dI}}}{\omega_{\mathrm{a}}}=\sqrt{1-\frac{4\omega_{1}%
^{2}}{\omega_{\mathrm{a}}^{2}}-\frac{\left(  2\gamma_{1}+\gamma_{2}\right)
\gamma_{2}}{\omega_{\mathrm{a}}^{2}}}\;. \label{omega_dI}%
\end{equation}
Using these notations Eq. (\ref{Lambda_1= V1}) becomes%
\begin{equation}
\frac{\Lambda_{1}}{\omega_{\mathrm{a}}}=\frac{8g_{\mathrm{a}}^{2}\omega_{1}%
}{\omega_{\mathrm{a}}^{3}}\frac{ip_{+}^{\prime\prime}\frac{\Delta
_{\mathrm{pL}}}{\omega_{\mathrm{a}}}+p_{+}^{\prime}\left(  1+i\frac{\gamma
_{2}}{\omega_{\mathrm{a}}}\right)  }{\frac{D_{\mathrm{L}}}{\omega_{\mathrm{a}%
}^{3}}}\;, \label{Lambda_1= V2}%
\end{equation}
where $p_{+}^{\prime}$ ($p_{+}^{\prime\prime}$) is the real (imaginary) part
of $p_{+}$, i.e.%
\begin{align}
p_{+}^{\prime}  &  =\frac{p_{+}+p_{+}^{\ast}}{2}\;,\\
p_{+}^{\prime\prime}  &  =\frac{p_{+}-p_{+}^{\ast}}{2i}\;.
\end{align}

To second order in $g_{\mathrm{a}}$ the term $\Lambda_{1}$ [see Eq.
(\ref{Lambda_1= V2})] can be calculated by evaluating the fixed point value of
$p_{+}$ to zeroth order in $g_{\mathrm{a}}$, which is done by solving the set
of equations $\Theta_{\mathrm{a}}=0$, $\Theta_{+}=0$ and $\Theta_{z}=0$ for
the case $g_{\mathrm{a}}=0$. The steady state values of the variables $a$,
$p_{+}$ and $p_{z}$ are found to be given by $a_{0}=0$,%
\begin{align}
p_{+0}  &  =\frac{\frac{\omega_{1}}{\gamma_{2}}\left(  -\frac{\Delta
_{\mathrm{pL}}}{\gamma_{2}}-i\right)  p_{0}}{1+\frac{\Delta_{\mathrm{pL}}^{2}%
}{\gamma_{2}^{2}}+\frac{4\omega_{1}^{2}}{\gamma_{1}\gamma_{2}}}\;,
\label{p_+0}\\
p_{z0}  &  =\frac{\left(  1+\frac{\Delta_{\mathrm{pL}}^{2}}{\gamma_{2}^{2}%
}\right)  p_{0}}{1+\frac{\Delta_{\mathrm{pL}}^{2}}{\gamma_{2}^{2}}%
+\frac{4\omega_{1}^{2}}{\gamma_{1}\gamma_{2}}}\;, \label{p_z0}%
\end{align}
respectively. For the case where $\gamma_{1},\gamma_{2}\ll\omega_{\mathrm{a}}$
Eqs. (\ref{omega_dR}) and (\ref{omega_dI}) become%
\begin{equation}
\frac{\omega_{\mathrm{dR}}}{\omega_{\mathrm{a}}}=\sqrt{1+\frac{2\gamma_{2}%
}{\gamma_{1}}\left(  1-\frac{2\omega_{1}^{2}}{\omega_{\mathrm{a}}^{2}}\right)
}\;, \label{omega_dR small gammas}%
\end{equation}
and%
\begin{equation}
\frac{\omega_{\mathrm{dI}}}{\omega_{\mathrm{a}}}=\sqrt{1-\frac{4\omega_{1}%
^{2}}{\omega_{\mathrm{a}}^{2}}}\;. \label{omega_dI small gammas}%
\end{equation}
With the help of Eqs. (\ref{D_a/omega_C^3}), (\ref{Lambda_1= V2}),
(\ref{omega_dR small gammas}) and (\ref{omega_dI small gammas}) one obtains
for this case%
\begin{equation}
\frac{\Lambda_{1}}{\omega_{\mathrm{a}}}=-\frac{\frac{\frac{8g_{\mathrm{a}}%
^{2}\omega_{1}^{2}}{\omega_{\mathrm{a}}^{3}\gamma_{2}}\frac{\Delta
_{\mathrm{pL}}}{\gamma_{2}}\left(  1+\frac{2i\gamma_{2}}{\omega_{\mathrm{a}}%
}\right)  }{1+\frac{\Delta_{\mathrm{pL}}^{2}}{\gamma_{2}^{2}}+\frac
{4\omega_{1}^{2}}{\gamma_{1}\gamma_{2}}}p_{0}}{\frac{\gamma_{1}}%
{\omega_{\mathrm{a}}}\left(  \frac{\omega_{\mathrm{R}}^{2}+\eta\omega
_{\mathrm{a}}^{2}}{\omega_{\mathrm{a}}^{2}}-1\right)  -i\left(  \frac
{\omega_{\mathrm{R}}^{2}}{\omega_{\mathrm{a}}^{2}}-1\right)  }\;,
\label{Lambda_1= V3}%
\end{equation}
where%
\begin{equation}
\eta=\frac{2\gamma_{2}}{\gamma_{1}}\left[  \left(  1-\frac{\gamma_{1}}%
{\gamma_{2}}\right)  \frac{2\omega_{1}^{2}}{\omega_{\mathrm{a}}^{2}}-1\right]
\;, \label{eta=}%
\end{equation}
and where $\omega_{\mathrm{R}}=\sqrt{4\omega_{1}^{2}+\Delta_{\mathrm{pL}}^{2}%
}$ is the Rabi frequency of the driven spins.

\section{Intermode coupling}

In general, Eq. (\ref{lambda_1=}) can be employed for calculating the
eigenvalue of a cavity mode that is weakly coupled to any given ancila system.
In the previous section the ancila system under consideration was an ensemble
of driven spins, whereas in the current section the ancila system is taken to
be the driven spiral mode. In general, the second order term $-g^{2}\left(
VR\left(  \omega_{\mathrm{a}}\right)  V\right)  _{11}$ in Eq. (\ref{lambda_1=}%
) can be calculated by evaluating the steady state response of the ancilla
system to small monochromatic oscillations of the cavity mode at its own
resonance frequency. Substituting the steady state solution into the equation
of motion of the cavity mode gives its eigenvalue. This approach will be
employed in this section.

The Hamiltonian of the two-mode cavity closed system is taken to be given by%
\begin{align}
\hbar^{-1}\mathcal{H}_{\mathrm{ab}}  &  =\omega_{\mathrm{a}}\left(
A_{\mathrm{a}}^{\dag}A_{\mathrm{a}}+\frac{1}{2}\right)  +\omega_{\mathrm{b}%
}\left(  A_{\mathrm{b}}^{\dag}A_{\mathrm{b}}+\frac{1}{2}\right) \nonumber\\
&  +K\left(  A_{\mathrm{a}}+A_{\mathrm{a}}^{\dag}\right)  \left(
A_{\mathrm{b}}+A_{\mathrm{b}}^{\dag}\right)  ^{2}\ ,\nonumber\\
&  \label{H_ab}%
\end{align}
where $\omega_{\mathrm{a}}$ and $A_{\mathrm{a}}$ ($\omega_{\mathrm{b}}$ and
$A_{\mathrm{b}}$) are the angular frequency and the annihilation operator,
respectively, of the omega (spiral) resonator, and $K$ is the intermode
coupling coefficient. The Heisenberg equations of motion are given by [see Eq.
(\ref{Heisenberg})]%
\begin{align}
\frac{\mathrm{d}A_{\mathrm{a}}}{\mathrm{d}t}+i\omega_{\mathrm{a}}%
A_{\mathrm{a}}+iK\left(  A_{\mathrm{b}}+A_{\mathrm{b}}^{\dag}\right)  ^{2}  &
=0\;,\\
\frac{\mathrm{d}A_{\mathrm{b}}}{\mathrm{d}t}+i\omega_{\mathrm{b}}%
A_{\mathrm{b}}+2iK\left(  A_{\mathrm{a}}+A_{\mathrm{a}}^{\dag}\right)  \left(
A_{\mathrm{b}}+A_{\mathrm{b}}^{\dag}\right)   &  =0\;.
\end{align}
Adding damping and driving leads to%
\begin{equation}
\frac{\mathrm{d}A_{\mathrm{a}}}{\mathrm{d}t}+\left(  i\omega_{\mathrm{a}%
}+\gamma_{\mathrm{a}}\right)  A_{\mathrm{a}}+iK\left(  A_{\mathrm{b}%
}+A_{\mathrm{b}}^{\dag}\right)  ^{2}=F_{\mathrm{a}}\;, \label{A_1 dot}%
\end{equation}
and%
\begin{align}
&  \frac{\mathrm{d}A_{\mathrm{b}}}{\mathrm{d}t}+\left(  i\omega_{\mathrm{b}%
}+\gamma_{\mathrm{b}}\right)  A_{\mathrm{b}}+2iK\left(  A_{\mathrm{a}%
}+A_{\mathrm{a}}^{\dag}\right)  \left(  A_{\mathrm{b}}+A_{\mathrm{b}}^{\dag
}\right) \nonumber\\
&  =F_{\mathrm{bf}}e^{-i\left(  \omega_{\mathrm{b}}+\omega_{\mathrm{D}%
}\right)  t}+F_{\mathrm{b}}\;,\nonumber\\
&  \label{A_2 dot}%
\end{align}
where both noise terms $F_{\mathrm{a}}$ and $F_{\mathrm{b}}$ have a vanishing
expectation value. Averaging yields%
\begin{equation}
\frac{\mathrm{d}\mathcal{A}_{\mathrm{a}}}{\mathrm{d}t}+\left(  i\omega
_{\mathrm{a}}+\gamma_{\mathrm{a}}\right)  \mathcal{A}_{\mathrm{a}}+iK\left(
\mathcal{A}_{\mathrm{b}}+\mathcal{A}_{\mathrm{b}}^{\ast}\right)  ^{2}=0\;,
\label{A_1 dot Av}%
\end{equation}
and%
\begin{align}
&  \frac{\mathrm{d}\mathcal{A}_{\mathrm{b}}}{\mathrm{d}t}+\left(
i\omega_{\mathrm{b}}+\gamma_{\mathrm{b}}\right)  \mathcal{A}_{\mathrm{b}%
}+S_{\mathrm{b}1}+S_{\mathrm{b}2}\nonumber\\
&  =F_{\mathrm{bf}}e^{-i\left(  \omega_{\mathrm{b}}+\omega_{\mathrm{D}%
}\right)  t}\;.\nonumber\\
&  \label{A_2 dot Av}%
\end{align}
where%
\begin{align}
\left\langle A_{\mathrm{a}}\right\rangle  &  =\mathcal{A}_{\mathrm{a}%
}=a_{\mathrm{a}}e^{-i\omega_{\mathrm{a}}t}\;,\label{A_n Av}\\
\left\langle A_{\mathrm{b}}\right\rangle  &  =\mathcal{A}_{\mathrm{b}%
}=a_{\mathrm{b}}e^{-i\omega_{\mathrm{b}}t}\;,
\end{align}
and where%
\begin{align}
S_{\mathrm{b}1}  &  =2iK\left(  \mathcal{A}_{\mathrm{a}}+\mathcal{A}%
_{\mathrm{a}}^{\ast}\right)  \mathcal{A}_{\mathrm{b}}\;,\label{S_b1}\\
S_{\mathrm{b}2}  &  =2iK\left(  \mathcal{A}_{\mathrm{a}}+\mathcal{A}%
_{\mathrm{a}}^{\ast}\right)  \mathcal{A}_{\mathrm{b}}^{\ast}\;. \label{S_b2}%
\end{align}
In the subsections below the effect of the terms $S_{\mathrm{b}1}$ and
$S_{\mathrm{b}2}$ is separately evaluated.

\subsection{The effect of the $S_{\mathrm{b}1}$ term}

When the term $S_{\mathrm{b}2}$ is disregarded Eq. (\ref{A_2 dot Av}) becomes%
\begin{equation}
\frac{\mathrm{d}C_{\mathrm{b}}}{\mathrm{d}t}+\left(  i\Omega_{\mathrm{b}%
}+\gamma_{\mathrm{b}}\right)  C_{\mathrm{b}}=F_{\mathrm{bf}}\;,
\label{C_b eom}%
\end{equation}
where%
\begin{equation}
\Omega_{\mathrm{b}}=-\omega_{\mathrm{D}}+2K\left(  \mathcal{A}_{\mathrm{a}%
}+\mathcal{A}_{\mathrm{a}}^{\ast}\right)  \;,
\end{equation}
and where%
\begin{equation}
\mathcal{A}_{\mathrm{b}}=C_{\mathrm{b}}e^{-i\left(  \omega_{\mathrm{b}}%
+\omega_{\mathrm{D}}\right)  t}\;.
\end{equation}

By employing the notation%
\begin{equation}
C_{\mathrm{b}}=C_{\mathrm{b}0}+c_{\mathrm{b}}\;,
\end{equation}
where%
\begin{equation}
C_{\mathrm{b}0}=\frac{F_{\mathrm{bf}}}{-i\omega_{\mathrm{D}}+\gamma
_{\mathrm{b}}}\;, \label{C_20}%
\end{equation}
one obtains in the limit of small $K$%
\begin{equation}
\frac{\mathrm{d}c_{\mathrm{b}}}{\mathrm{d}t}+\left(  -i\omega_{\mathrm{D}%
}+\gamma_{\mathrm{b}}\right)  c_{\mathrm{b}}=-2iK\left(  \mathcal{A}%
_{\mathrm{a}}+\mathcal{A}_{\mathrm{a}}^{\ast}\right)  C_{\mathrm{b}0}\;.
\end{equation}
Let $\mathcal{A}_{\mathrm{a}}=a_{\mathrm{a}}e^{-i\omega_{\mathrm{a}}t}$ [see
Eq. (\ref{A_n Av})], and assume that $a_{\mathrm{a}}$ is constant. The steady
state solution reads%
\begin{equation}
c_{\mathrm{b}}=\frac{2iKC_{\mathrm{b}0}\mathcal{A}_{\mathrm{a}}}{i\left(
\omega_{\mathrm{D}}+\omega_{\mathrm{a}}\right)  -\gamma_{\mathrm{b}}}%
+\frac{2iKC_{\mathrm{b}0}\mathcal{A}_{\mathrm{a}}^{\ast}}{i\left(
\omega_{\mathrm{D}}-\omega_{\mathrm{a}}\right)  -\gamma_{\mathrm{b}}}\;.
\end{equation}
When only terms proportional to $\mathcal{A}_{\mathrm{a}}$ are kept, one finds
the coupling term in Eq. (\ref{A_1 dot Av}) can be expressed as%
\begin{align}
&  iK\left(  \mathcal{A}_{\mathrm{b}}+\mathcal{A}_{\mathrm{b}}^{\ast}\right)
^{2}\nonumber\\
&  \simeq\frac{4iK^{2}\left\vert C_{\mathrm{b}0}\right\vert ^{2}%
\omega_{\mathrm{D}}\mathcal{A}_{\mathrm{a}}}{\left[  i\left(  \omega
_{\mathrm{a}}-\omega_{\mathrm{D}}\right)  -\gamma_{\mathrm{b}}\right]  \left[
i\left(  \omega_{\mathrm{a}}+\omega_{\mathrm{D}}\right)  -\gamma_{\mathrm{b}%
}\right]  }\;.\nonumber\\
&  \label{CT1}%
\end{align}

\subsection{The effect of the $S_{\mathrm{b}2}$ term}

For this case the term $S_{\mathrm{b}1}$ in Eq. (\ref{A_2 dot Av}) is
disregarded. Furthermore, the counter rotating term proportional to
$\mathcal{A}_{\mathrm{a}}^{\ast}\mathcal{A}_{\mathrm{b}}^{\ast}$ is
disregarded as well [see Eq. (\ref{S_b2})]. For this case Eq.
(\ref{A_2 dot Av}) becomes%
\begin{equation}
\frac{\mathrm{d}a_{\mathrm{b}}}{\mathrm{d}t}+\gamma_{\mathrm{b}}a_{\mathrm{b}%
}+2iKa_{\mathrm{a}}a_{\mathrm{b}}^{\ast}e^{i\omega_{\mathrm{s}}t}%
=F_{\mathrm{bf}}e^{-i\omega_{\mathrm{D}}t}\;, \label{a_2 dot}%
\end{equation}
where%
\begin{equation}
\omega_{\mathrm{s}}=2\omega_{\mathrm{b}}-\omega_{\mathrm{a}}\;.
\end{equation}
Consider a solution of Eq. (\ref{a_2 dot}) having the form \cite{Lifshitz2008}%
\begin{equation}
a_{\mathrm{b}}=\alpha e^{-i\omega_{\mathrm{D}}t}+\beta e^{i\left(
\omega_{\mathrm{s}}+\omega_{\mathrm{D}}\right)  t}\;. \label{a_2 alpha beta}%
\end{equation}
Substituting the solution into Eq. (\ref{a_2 dot}) and assuming that $\alpha$,
$\beta$ and $a_{\mathrm{a}}$ are all constants lead to%
\begin{equation}
\left(  -i\omega_{\mathrm{D}}+\gamma_{\mathrm{b}}\right)  \alpha
+2iKa_{\mathrm{a}}\beta^{\ast}=F_{\mathrm{bf}}\;,
\end{equation}
and%
\begin{equation}
\left[  i\left(  \omega_{\mathrm{s}}+\omega_{\mathrm{D}}\right)
+\gamma_{\mathrm{b}}\right]  \beta+2iKa_{\mathrm{a}}\alpha^{\ast}=0\;,
\end{equation}
thus%
\begin{equation}
\alpha=\frac{F_{\mathrm{bf}}}{-i\omega_{\mathrm{D}}+\gamma_{\mathrm{b}}%
-\frac{4K^{2}\left\vert a_{\mathrm{a}}\right\vert ^{2}}{-i\left(
\omega_{\mathrm{s}}+\omega_{\mathrm{D}}\right)  +\gamma_{\mathrm{b}}}}\;,
\label{alpha=}%
\end{equation}
and%
\begin{equation}
\beta=\frac{-2iKa_{\mathrm{a}}\alpha^{\ast}}{i\left(  \omega_{\mathrm{s}%
}+\omega_{\mathrm{D}}\right)  +\gamma_{\mathrm{b}}}\;. \label{beta=}%
\end{equation}

The steady state solution (\ref{a_2 alpha beta}) can be used to express the
coupling term $iK\left(  \mathcal{A}_{\mathrm{b}}+\mathcal{A}_{\mathrm{b}%
}^{\ast}\right)  ^{2}$ in Eq. (\ref{A_1 dot Av}) in terms of $\mathcal{A}%
_{\mathrm{a}}$. To that end $\mathcal{A}_{\mathrm{b}}$ is expressed as [see
Eqs. (\ref{A_n Av}), (\ref{a_2 alpha beta}) and (\ref{beta=})]%
\begin{align}
\mathcal{A}_{\mathrm{b}}  &  =\alpha e^{-i\left(  \omega_{\mathrm{b}}%
+\omega_{\mathrm{D}}\right)  t}+\beta e^{i\left(  \omega_{\mathrm{b}}%
+\omega_{\mathrm{D}}-\omega_{\mathrm{a}}\right)  t}\nonumber\\
&  =\alpha e^{-i\left(  \omega_{\mathrm{b}}+\omega_{\mathrm{D}}\right)
t}-\frac{2iK\alpha^{\ast}e^{i\left(  \omega_{\mathrm{b}}+\omega_{\mathrm{D}%
}\right)  t}}{i\left(  \omega_{\mathrm{s}}+\omega_{\mathrm{D}}\right)
+\gamma_{\mathrm{b}}}\mathcal{A}_{\mathrm{a}}\;.\nonumber\\
&
\end{align}
When only terms proportional to $\mathcal{A}_{\mathrm{a}}$ are kept, the
following approximation is employed [see Eq. (\ref{alpha=})]%
\begin{equation}
\alpha\simeq\frac{F_{\mathrm{bf}}}{-i\omega_{\mathrm{D}}+\gamma_{\mathrm{b}}%
}\;,
\end{equation}
and it is assumed that $\left\vert \omega_{\mathrm{D}}\right\vert
\ll\left\vert \omega_{\mathrm{s}}\right\vert $ for evaluating $\beta$ [see Eq.
(\ref{beta=})] the coupling term in Eq. (\ref{A_1 dot Av}) becomes%
\begin{align}
iK\left(  \mathcal{A}_{\mathrm{b}}+\mathcal{A}_{\mathrm{b}}^{\ast}\right)
^{2}  &  \simeq-\frac{4K^{2}\left\vert \alpha\right\vert ^{2}\mathcal{A}%
_{\mathrm{a}}}{i\left(  \omega_{\mathrm{s}}+\omega_{\mathrm{D}}\right)
+\gamma_{\mathrm{b}}}\nonumber\\
&  \simeq-\frac{4K^{2}\left\vert F_{\mathrm{bf}}\right\vert ^{2}%
\mathcal{A}_{\mathrm{a}}}{\left(  i\omega_{\mathrm{s}}+\gamma_{\mathrm{b}%
}\right)  \left(  \omega_{\mathrm{D}}^{2}+\gamma_{\mathrm{b}}^{2}\right)
}\;.\nonumber\\
&
\end{align}
When $\gamma_{\mathrm{b}}\ll\omega_{\mathrm{s}}$ one has%
\begin{equation}
iK\left(  \mathcal{A}_{\mathrm{b}}+\mathcal{A}_{\mathrm{b}}^{\ast}\right)
^{2}\simeq\frac{4K^{2}\left\vert F_{\mathrm{bf}}\right\vert ^{2}\left(
i\omega_{\mathrm{s}}-\gamma_{\mathrm{b}}\right)  \mathcal{A}_{\mathrm{a}}%
}{\omega_{\mathrm{s}}^{2}\left(  \omega_{\mathrm{D}}^{2}+\gamma_{\mathrm{b}%
}^{2}\right)  }\;. \label{CT2}%
\end{equation}

\newpage
\bibliographystyle{ieee}
\bibliography{acompat,Eyal_Bib}

\end{document}